**Countries across the world use more land for golf courses than wind or solar energy**

Jann Michael Weinand[1], Tristan Pelser[1,2], Max Kleinebrahm[3], Detlef Stolten[1,2]

[1]Institute of Climate and Energy Systems (ICE) – Jülich Systems Analysis (ICE-2), Forschungszentrum Jülich GmbH, Jülich,Germany

[2]RWTH Aachen University, Chair for Fuel Cells, Faculty of Mechanical Engineering, Aachen, Germany

[3]Institute for Industrial Production (IIP), Karlsruhe Institute for Technology, Karlsruhe, Germany

Corresponding author: Jann Michael Weinand, j.weinand@fz-juelich.de, +49 175 4985402

**Abstract**

Land use is a critical factor in the siting of renewable energy facilities and is often scrutinized due to perceived conflicts with other land demands. Meanwhile, substantial areas are devoted to activities such as golf, which are accessible to only a select few and have a significant land and environmental footprint. Our study shows that in countries such as the United States and the United Kingdom, far more land is allocated to golf courses than to renewable energy facilities. Areas equivalent to those currently used for golf could support the installation of up to 842 GW of solar and 659 GW of wind capacity in the top ten countries with the most golf courses. In many of these countries, this potential exceeds both current installed capacity and medium-term projections. These findings underscore the untapped potential of rethinking land use priorities to accelerate the transition to renewable energy.

**Introduction**

The increasing urgency of addressing the global land squeeze and climate change requires a critical examination of land use priorities[1,2]. Golf courses, for example, cover large areas around the world[3,4], often in regions where land availability is limited. This has led to discussions about converting the courses to housing or public parkland, e.g., in Canada[5], Australia[6] and the UK[7–9]. In this paper, we argue that golf courses also provide a notable contrast to the growing demand for renewable energy infrastructure. A recent (non-peer-reviewed) analysis estimated that the total area of golf courses in Germany exceeds that of renewable energy facilities[4].

Golf courses typically require extensive turf maintenance, including significant water use and chemical treatments, resulting in a significant environmental impact[10–12]. In comparison, renewable energy installations such as solar farms and wind turbines offer a sustainable use of land and directly contribute to the reduction of greenhouse gas emissions. Utility-scale solar farms require approximately 0.01 km² per megawatt (MW),[13] while wind farms require around 0.12 km² per MW,[14] although only a small fraction of this land is directly impacted by turbines and infrastructure[15]. In addition, built-up land such as golf courses is typically not included in renewable energy potential analyses, further highlighting the importance of reevaluating land use options.

Furthermore, the cost of renewable energy to decarbonize energy systems has been falling for years[16,17], but realizing its high economic potential[18] is often hampered by social opposition[19,20], land use trade-offs[21], and siting regulations[22]. While golf courses could provide recreational and aesthetic value, their extensive land use invites a comparison with the pressing need for sustainable energy solutions. Evaluating the potential for converting golf course land into renewable energy sites offers a strategic perspective on optimizing land use for broader environmental and societal benefits.

**Results**

**More area for golf courses than renewables**

There are about 38,400 golf courses in the world, 80% of which are located in the top ten countries with most courses (see **Figure 1a**). With over 16,000 courses, the United States of America tops this list (see **Figure 1b**), followed by the United Kingdom (around 3,100) and Japan (around 2,700). With an average of approx. 0.8 km², the individual golf courses in China are the largest, with second-placed Japan, at 0.5 km², well behind and closer to the other countries. Moreover, golf courses occupy a considerable proportion of the land area within individual nations. In the United Kingdom, this figure reaches 0.49%, with South Korea and Japan following at 0.42% and 0.37%, respectively.

Although the data collection was challenging, it is possible to demonstrate at least for Europe that the number of golf courses has increased significantly since 1985 (see **Figure 1c**). Despite the accelerated expansion of wind turbine and solar photovoltaics (PV) capacity in recent years, the individual technologies have not been able to reach the requisite area for golf course development in the top ten countries (except for China). Even when assuming a figure of 0.015 km² per MW, it is evident that the area dedicated to golf courses is still significantly larger in some countries than that allocated to utility-scale PV. To illustrate, under this assumption, there would be approximately 16 times more space for golf courses in Canada, six times more in the United Kingdom, or four times more in the United States of America. Germany also has a golf course area that is 1.25 times larger than the area used for utility-scale PV. This corroborates the premise of this article, which is based on the (non-peer-reviewed) example of land area for golf courses versus utility-scale PV in Germany.

The number of golf courses is also expected to increase in the future. More than 500 golf courses are currently planned or under construction in 88 countries around the world. The majority of these, 124, are being built in the United States, but Vietnam is also experiencing an upswing with 51 new courses, ahead of the United Kingdom in third place (27).[23]

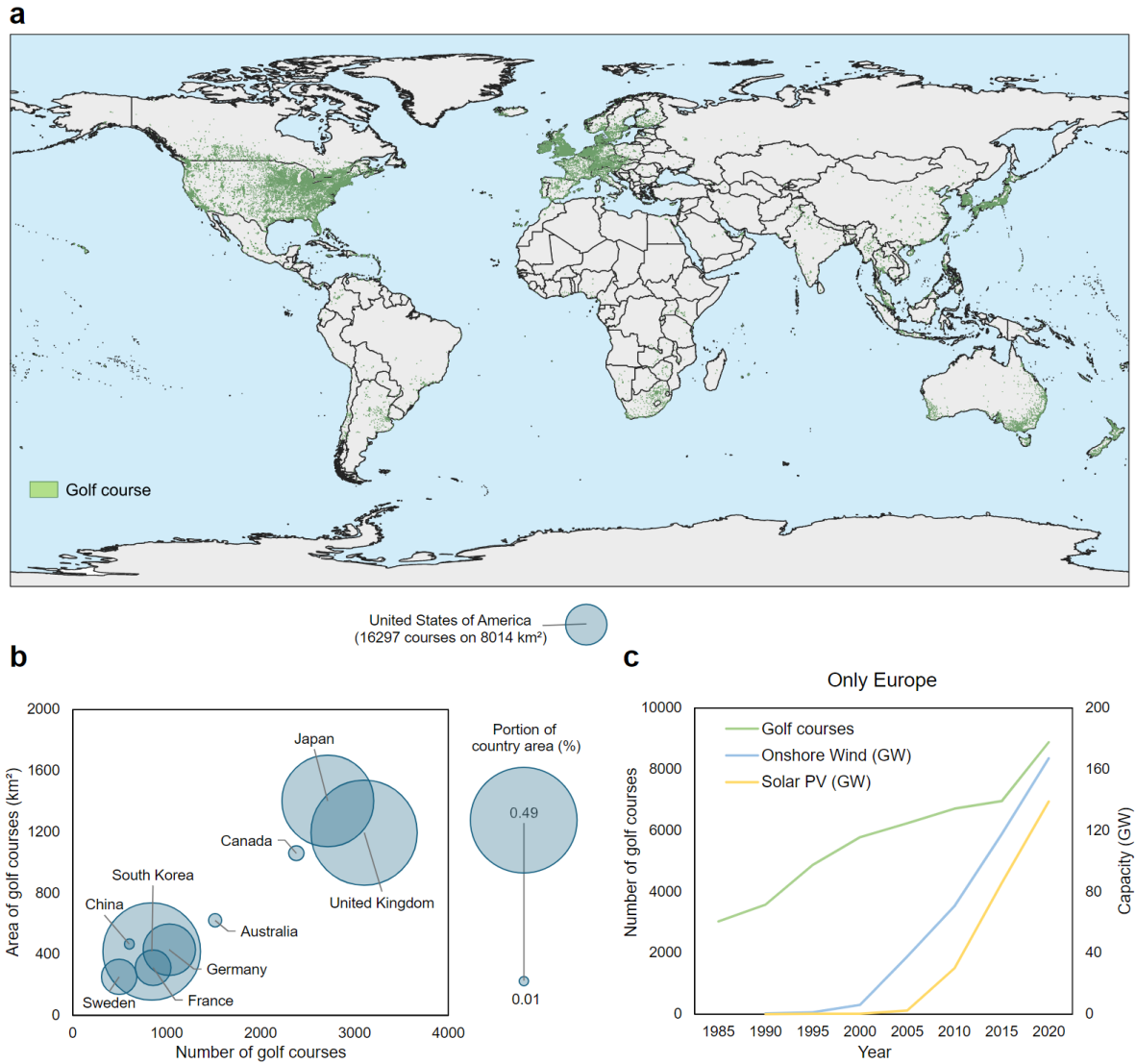

**Figure 1: Global land-use of golf courses.**
*a. The locations of 38,427 golf courses worldwide, obtained from OpenStreetMap[24]. b. Area and number of courses for the top ten countries with the most courses, as well as their portion of the country's land area. The United States of America is shown outside the graph to better illustrate the differences between the other countries. c. Temporal development of the number of golf courses[25–27] and the capacity[28–30] of onshore wind and solar PV in Europe.*

**Renewable energy potentials on golf courses**

The top 10 countries could install between 281 to 842 GW of utility-scale PV on 25% to 75% of the golf course land area (see **Figure 2**). The capacity for 75% coverage is slightly higher than the currently installed capacity in the countries[31–35], which totals 646 GW. However, the potential capacity of 842 GW is significantly higher than the currently installed capacity of 257 GW if China is excluded. In fact, the current capacity (excluding China) would even be exceeded if only 25% coverage were assumed on the golf courses. In the countries excluding China, even the capacity forecasts[36] up to 2028 (496 GW) could be achieved with 50% coverage (543 GW) through installations on golf course areas. Although this applies in total, it does not apply to Germany, South Korea and France, apart from China, where the area on golf courses would not be sufficient to meet the capacity projections.

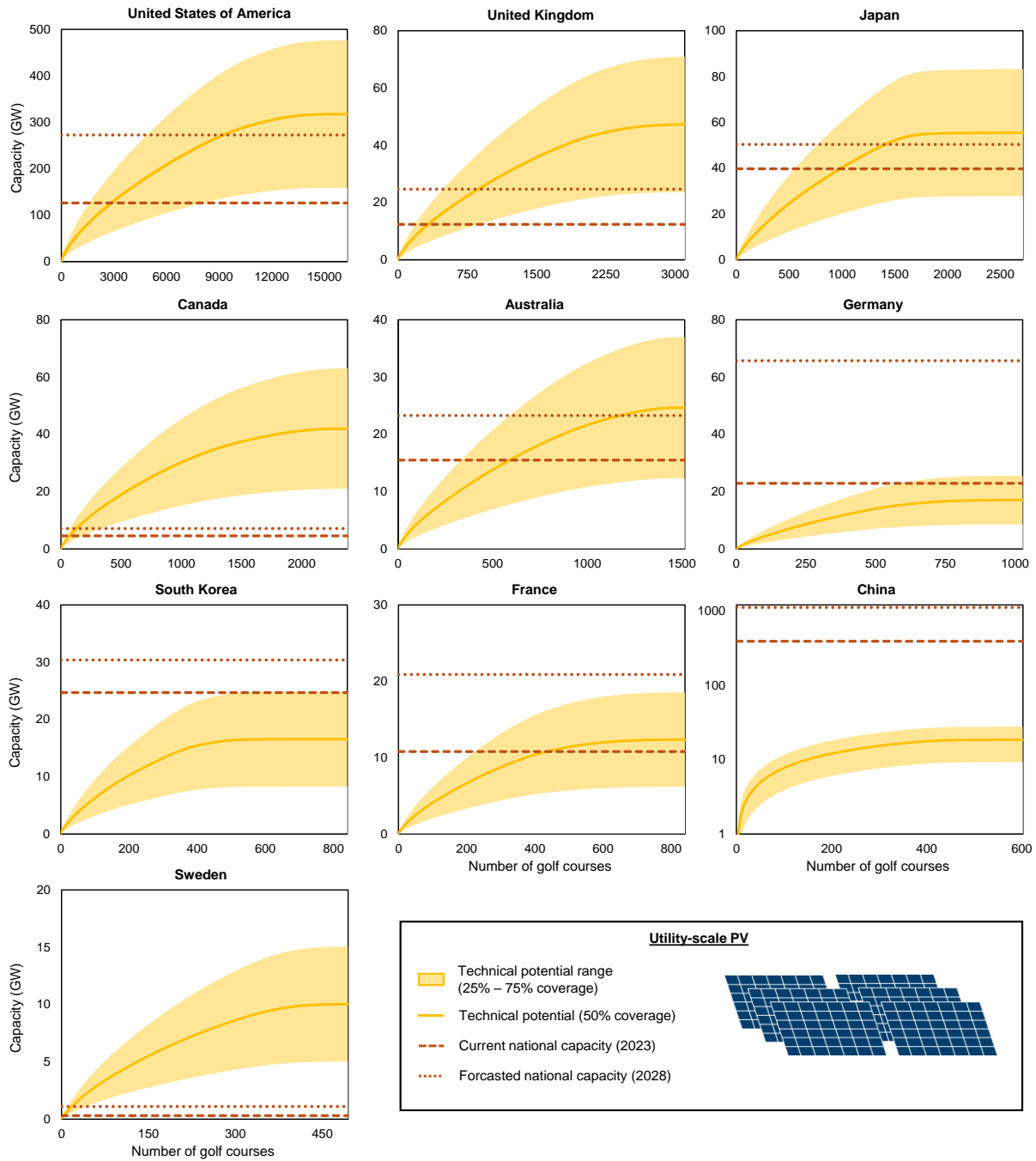

*Figure 2: Technical utility-scale PV potential on golf courses in the top ten countries with the most golf courses in the world, as well as current installed and projected national capacity.*
*The golf courses are shown in descending order of size (x-axis). Current figures for 2023 are based on statistical data[31–35] and projected figures for 2028 are based on the International Energy Agency's (IEA) "Main Case" scenario[36]. Given that the existing and future capacities of China are significantly higher than the potential on golf courses, the y-axis is presented in a logarithmic format for this country.*

Depending on turbine spacing (see Methods), the top 10 countries could alternatively install between 174 and 659 GW of onshore wind across the entire golf course area (see **Figure 3**). At 702 GW, the current capacities[31–35] of these countries are above the maximum potential, although they are within the range of the potential at 298 GW if China is again excluded. The capacity forecast[36] of 443 GW in 2028 (excluding China) can also be achieved by golf courses overall, but not in Germany, France, China and Sweden.

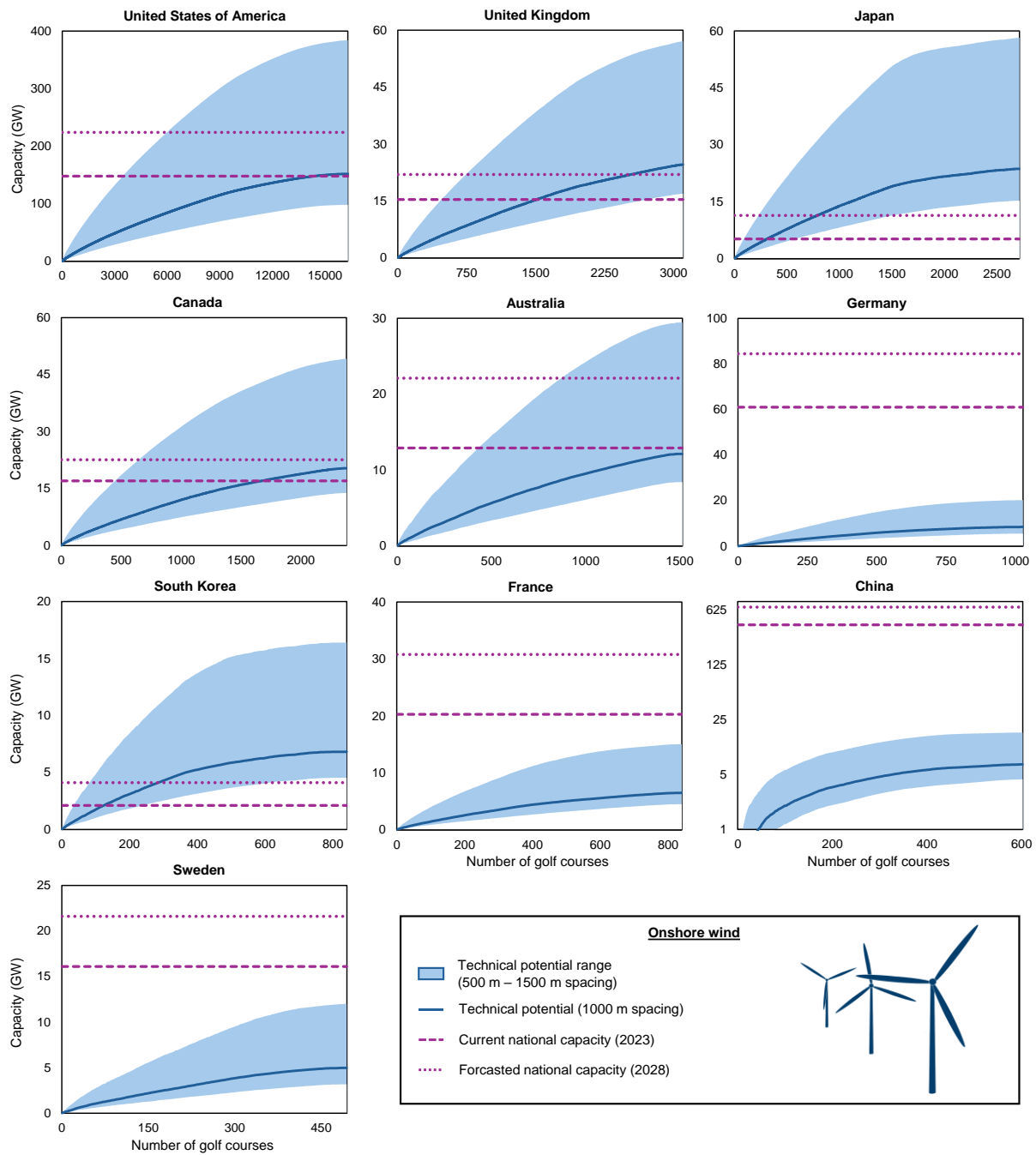

*Figure 3: Technical onshore wind potential on golf courses in the top ten countries with the most golf courses in the world, as well as current installed and projected national capacity.*
*The golf courses are shown in descending order of size (x-axis). Current figures for 2023 are based on statistical data[31–35] and projected figures for 2028 are based on the International Energy Agency's (IEA) "Main Case" scenario[36]. Given that the existing and future capacities of China are significantly higher than the potential on golf courses, the y-axis is presented in a logarithmic format for this country.*

When looking at individual countries, it is noticeable that the golf course areas in China, Germany and France in particular would be far from sufficient to achieve the ambitious expansion targets for renewable energy facilities. In addition, some countries prioritize the expansion of onshore wind (Canada and Sweden) and others the expansion of utility-scale PV (Japan and South Korea). The different prioritization has various reasons such as a strong solar lobby in Japan[37,38] or limited policy incentives for solar PV in Sweden[39]. As a result, in these countries except for Japan, the area on the golf courses would be sufficient for the less prioritized technology but not for the prioritized one. By contrast, in countries such as the United States of America, the United Kingdom, Japan or Australia, the capacity forecasts up to 2028 and beyond could comfortably be achieved with an area equivalent to the golf course land.

**Discussion**

This study does not advocate that golf courses should be completely replaced but is intended to provide a perspective on how much renewable energy systems could be installed on a comparable area. It is also important to note that the energy use on golf courses in the United States has decreased since the early 2000s, in particular due to the reduction of the maintained turfgrass acreage and the increased use of clean energy sources[40,41]. Furthermore, some golf courses are located in the immediate vicinity of urban areas and would therefore not be entirely suitable for the installation of wind turbines. Two well-known examples are The Old Course at St. Andrews, Scotland, which is partially surrounded by buildings, and Augusta National in Georgia, United States, which is completely surrounded by the urban area of the city of Augusta.

Conversely, golf courses, which often occupy vast areas of land and are accessible only through exclusive membership[6], highlight significant land use conflicts. The concept of energy justice argues that the benefits and burdens of energy supply and land use should be distributed equitably across society[42,43]. In this context, these exclusive golf courses could be repurposed for renewable energy projects that benefit the wider public. Importantly, this does not mean that golf must be entirely abandoned. For instance, in South Korea, screen golf—a popular alternative involving indoor simulators—offers a more accessible, convenient, and affordable way to enjoy the sport compared to traditional outdoor courses[44]. The feasibility of converting golf course land, even in rural areas, has been demonstrated in the past. In Japan, for example, in a rural area in the Hyogo Prefecture, an entire golf course was converted into a solar park with 260,000 solar panels (125 GWh).[45]

Furthermore, the use of renewable energy would not necessarily prevent golf from continuing to be played on the courses. The deployment of wind turbines has the advantage over utility-scale PV of requiring less land for the actual installations (see **Figure 4**). In the United States, for example, depending on the spacing between the turbines, an average of only between 1.1 (1,500 m spacing) and 4.3 (500 m) turbines would be installed per golf course. The hybrid utilization of the golf courses would probably also be possible with up to 25% area coverage by solar modules and could be a possible strategy for the future.

In this article, we were able to show that in some countries, more land is used for the exclusive recreational sport of golf than for utility-scale PV, and that an area the size of the golf courses would be sufficient to meet the medium-term development targets for onshore wind or solar

PV in these countries. This provides a useful context for the land requirements of renewables, which are the subject of considerable debate. However, the long-term energy transition will require significantly more land for renewables than golf courses currently occupy. For example, a 100% clean electricity scenario for the United States of America in 2035 would require between 5,000–9,000 km² for onshore wind and 15,000–29,000 km² for utility-scale solar[46], which would account for up to 0.4% of the total country area. This would be up to 4.7 times more land than the golf courses currently account for in the United States, but still significantly less than the land for active oil and gas leases in 2020 (~105,000 km²)[46]. In addition, in Germany, the complete decarbonization of the energy system would require at least 6% of the country's land area (21,450 km²) for renewable energies. However, the 4% for utility-scale photovoltaics and 2% for onshore wind would also replace the 6.5% of land currently used for bioenergy crops[47].

**Methods**

**Golf course data.** The area, location and metadata for golf courses were obtained from OpenStreetMap using the following query:

*[out:json][timeout:99999];*

*nwr["leisure"="golf_course"]({{bbox}});*

*out geom;*

OpenStreetMap contains 38,427 golf courses, which is 1.1% less than the actual number. This is particularly evident in Sweden, Japan and Canada, where 25%, 12% and 6% of courses are missing, respectively. However, the two countries with the most golf courses, the United States (+1%) and the United Kingdom (+/- 0%), as well as most other countries, have very good coverage[26,48]. The area of the golf courses also largely matches, with a deviation of only 1% for the United States[3].

**Renewable energy potential assessment.**

The Renewable Energy Potentials Workflow Manager (REFLOW)[49] was used to develop and execute the software workflow for this analysis. REFLOW is a python-based workflow manager built on Luigi[50] and was developed in response to the fact that many assessments of renewable potential are not accurate or reproducible[18]. It facilitates transparency and reproducibility in renewable energy potential assessments by defining, managing, and running series of tasks relating to data acquisition, pre-processing, analysis and reporting. In this study, we employ REFLOW to manage the data workflow using data acquisition and pre-processing scripts, and then to interface with additional software for the land eligibility assessment (Geospatial Land Availability for Energy Systems (GLAES))[51].

For the wind assessment, we employ a single theoretical turbine model based on near-future designs[16] with a capacity of 5.5 MW, and rotor diameter of 135 meters. Since the golf courses are small areas, the low number of turbines that can be distributed on them reduces the need for staggered spacing to reduce wake effects. GLAES is employed to distribute turbines over the golf course polygons for three scenarios: high-, medium- and low-capacity density, equivalent to 1500 m, 1000 m and 500 m spacing between turbines (see **Figure 4**). For the solar PV assessment, we employ three scenarios based on the assumed percentage coverage of the golf courses by horizontal PV panels (75%, 50% and 25% coverage). This is to account for vegetation or water bodies that would mean that not all the space is available for the installations. We assume a capacity density of 79,2 MW/km$^2$,[52] such that the total installed capacity for each scenario is calculated by:

$$Installed\ capacity = Area_{golf\ course} \times 79{,}2 \frac{MW}{km^2} \times coverage$$

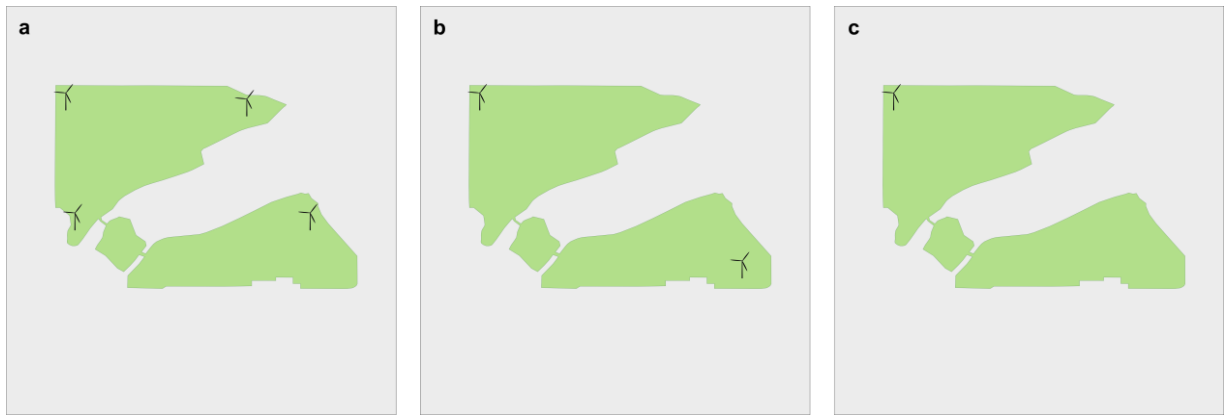

*Figure 4: Example wind turbine placement for a golf course in Nebraska, United States.*
*The panels show the placement for 500 m (a), 1000 m (b) and 1500 m (c) spacing between turbines.*

**Acknowledgements**

This work was supported by the Helmholtz Association under the program "Energy System Design".

**Data and code availability**

The ETHOS.REFLOW workflow manager for assessing the renewable energy potentials is available as open source on GitHub (https://github.com/FZJ-IEK3-VSA/ethos.REFLOW).

**Declaration of Generative AI and AI-assisted technologies in the writing process**

During the preparation of this work the authors used the tools "ChatGPT" and "DeepL" in order to check grammar and spelling in a few places, and to make minor improvements to readability and style. After using this tool, the authors reviewed and edited the content as needed, and take full responsibility for the content of the publication.

# References

1. Erb, K.-H., Matej, S., Haberl, H. & Gingrich, S. Sustainable land systems in the Anthropocene: Navigating the global land squeeze. *One Earth* **7,** 1170–1186; 10.1016/j.oneear.2024.06.011 (2024).

2. Gottdenker, N. L. & Chaves, L. F. Dispossession, displacement, and disease: The global land squeeze and infectious disease emergence. *One Earth* **7,** 1137–1141; 10.1016/j.oneear.2024.06.019 (2024).

3. Bloomberg. Here's How America Uses Its Land. Available at https://www.bloomberg.com/graphics/2018-us-land-use/?terminal=true (2018).

4. PV Magazine. Golf courses consume more land than PV in Germany. Available at https://www.pv-magazine.com/2024/05/23/golf-courses-consume-more-land-than-pv-in-germany/ (2024).

5. Couture, J., Millington, B. & Wilson, B. Who is the city for? Sports facilities and the case of Vancouver's public golf courses. *International Journal of Sport Policy and Politics* **15,** 45–62; 10.1080/19406940.2022.2161601 (2023).

6. Australian Broadcasting Corporation. Environmentalist argues golf courses should be 'fair game' in pursuit of more green public space. Available at https://www.abc.net.au/news/2023-10-29/should-sydney-use-golf-courses-for-public-parks/103032260 (2023).

7. The Guardian. Building houses on Britain's vast, exclusive golf courses makes sense for everyone – even golfers. Available at https://www.theguardian.com/commentisfree/2023/nov/28/building-houses-britain-golf-courses-makes-sense (2023).

8. British Broadcasting Corporation. How much of the UK is covered in golf course? Available at https://www.bbc.co.uk/news/magazine-24378868 (2013).

9. Financial Times. FT Factcheck: Do we use more land for golf courses than we do for homes? (2016).

10. Bekken, M. A. H. & Soldat, D. J. Estimated energy use and greenhouse gas emissions associated with golf course turfgrass maintenance in the Northern USA. *Intl Turfgrass Soc Res J* **14,** 58–75; 10.1002/its2.61 (2022).

11. Tidåker, P., Wesström, T. & Kätterer, T. Energy use and greenhouse gas emissions from turf management of two Swedish golf courses. *Urban Forestry & Urban Greening* **21,** 80–87; 10.1016/j.ufug.2016.11.009 (2017).

12. Wheeler, K. & Nauright, J. A Global Perspective on the Environmental Impact of Golf. *Sport in Society* **9,** 427–443; 10.1080/17430430600673449 (2006).

13. Bolinger, M. & Bolinger, G. Land Requirements for Utility-Scale PV: An Empirical Update on Power and Energy Density. *IEEE J. Photovoltaics* **12,** 589–594; 10.1109/JPHOTOV.2021.3136805 (2022).

14. Wachs, E. & Engel, B. Land use for United States power generation: A critical review of existing metrics with suggestions for going forward. *Renewable and Sustainable Energy Reviews* **143,** 110911; 10.1016/j.rser.2021.110911 (2021).

15. Harrison-Atlas, D., Lopez, A. & Lantz, E. Dynamic land use implications of rapidly expanding and evolving wind power deployment. *Environ. Res. Lett.* **17,** 44064; 10.1088/1748-9326/ac5f2c (2022).

16. Wiser, R. *et al.* Expert elicitation survey predicts 37% to 49% declines in wind energy costs by 2050. *Nat Energy* **6,** 555–565; 10.1038/s41560-021-00810-z (2021).

17. Haegel, N. M. *et al.* Photovoltaics at multi-terawatt scale: Waiting is not an option. *Science (New York, N.Y.)* **380,** 39–42; 10.1126/science.adf6957 (2023).

18. Pelser, T. *et al.* Reviewing accuracy & reproducibility of large-scale wind resource assessments. *Advances in Applied Energy* **13,** 100158; 10.1016/j.adapen.2023.100158 (2024).

19. Tsani, T., Weinand, J. M., Linßen, J. & Stolten, D. Quantifying social factors for onshore wind planning – A systematic review. *Renewable and Sustainable Energy Reviews* **203,** 114762; 10.1016/j.rser.2024.114762 (2024).

20. Weinand, J. M. *et al.* Historic drivers of onshore wind power siting and inevitable future trade-offs. *Environ. Res. Lett.* **17,** 74018; 10.1088/1748-9326/ac7603 (2022).

21. McKenna, R. *et al.* Exploring trade-offs between landscape impact, land use and resource quality for onshore variable renewable energy: an application to Great Britain. *Energy* **250,** 123754; 10.1016/j.energy.2022.123754 (2022).

22. Lopez, A. *et al.* Impact of siting ordinances on land availability for wind and solar development. *Nat Energy* **8,** 1034–1043; 10.1038/s41560-023-01319-3 (2023).

23. National Golf Foundation. Midyear Update: Worldwide Golf Course Development. Available at https://www.ngf.org/midyear-update-worldwide-golf-course-development/ (2024).

24. OpenStreetMap contributors. Golf course data retrieved from overpass-turbo.eu. Available at https://www.openstreetmap.org (2024).

25. R&A. Golf Around the World. Available at https://assets-us-01.kc-usercontent.com/c42c7bf4-dca7-00ea-4f2e-373223f80f76/50ff4344-b576-4e2e-a9e2-8411712954ac/2021%20Golf%20Around%20The%20World%20Fourth%20Edition.pdf (2021).

26. R&A. Global Golf Participation Report. Available at https://assets.randa.org/c42c7bf4-dca7-00ea-4f2e-373223f80f76/ed52a88d-f532-4d27-9606-ec94f8af9430/The%20R%26A_Global%20Golf%20Participation%20Report%202023.pdf (2023).

27. Statista. Number of official golf courses in Europe from 1985 to 2018. Available at https://www.statista.com/statistics/275309/number-of-golf-courses-in-europe/ (2019).

28. WindEurope. Wind energy in Europe 2020 Statistics and the outlook for 2021-2025. Available at https://windeurope.org/intelligence-platform/product/wind-energy-in-europe-2020-statistics-and-the-outlook-for-2021-2025/#:~:text=25%20February%202021-,Overview,on%20the%20onshore%20wind%20sector (2021).

29. eurostat. Electrical capacity for wind and solar photovoltaic power - statistics. Available at https://ec.europa.eu/eurostat/statistics-explained/index.php?title=Electrical_capacity_for_wind_and_solar_photovoltaic_power_-_statistics#Increasing_capacity_for_wind_and_solar_over_the_last_decades (2021).

30. Intersolar Europe. TREND PAPER FOR INTERSOLAR EUROPE: EU MARKET OUTLOOK FOR SOLAR POWER 2021-2025. Available at https://www.intersolar.de/media/doc/622b07fd3f2dd401c72b1754#:~:text=In%20total%2C%20the%20amount%20of,for%20almost%20half%20of%20this.&text=SolarPower%20Europe%20expects%20the%20PV,over%20the%20next%20four%20years. (2022).

31. IRENA. Wind energy. Available at https://www.irena.org/Energy-Transition/Technology/Wind-energy (2024).

32. PV Magazine. US utility-scale solar capacity additions hit 36.4 GW in 2023. Available at https://www.pv-magazine.com/2024/02/19/us-utility-scale-solar-capacity-additions-hit-36-4-gw-in-2023/ (2024).

33. Taiyang News. Canada Installed Over 400 MW New Solar PV Capacity In 2023. Available at https://taiyangnews.info/markets/canada-installed-over-400-mw-new-solar-pv-capacity-in-2023#:~:text=Utility%20scale%20PV%20added%20nearly,at%20the%20end%20of%202023. (2024).

34. Mercom. Australia adds 5.9 GW of renewable capacity to the Grid in 2023. Available at https://www.mercomindia.com/australia-adds-capacity-to-the-grid-in-2023 (2024).

35. Atlantic Council. China builds more utility-scale solar as competition with coal ramps up. Available at https://www.atlanticcouncil.org/blogs/energysource/china-builds-more-utility-scale-solar-as-competition-with-coal-ramps-up/ (2024).

36. IEA. Renewable Energy Progress Tracker. Available at https://www.iea.org/data-and-statistics/data-tools/renewable-energy-progress-tracker (2024).

37. Li, A., Xu, Y. & Shiroyama, H. Solar lobby and energy transition in Japan. *Energy Policy* **134,** 110950; 10.1016/j.enpol.2019.110950 (2019).

38. Cherp, A., Vinichenko, V., Jewell, J., Suzuki, M. & Antal, M. Comparing electricity transitions: A historical analysis of nuclear, wind and solar power in Germany and Japan. *Energy Policy* **101,** 612–628; 10.1016/j.enpol.2016.10.044 (2017).

39. Lindahl, J., Lingfors, D., Elmqvist, Å. & Mignon, I. Economic analysis of the early market of centralized photovoltaic parks in Sweden. *Renewable Energy* **185,** 1192–1208; 10.1016/j.renene.2021.12.081 (2022).

40. Shaddox, T. W., Unruh, J. B., Johnson, M. E., Brown, C. D. & Stacey, G. Land-use and Energy Practices on US Golf Courses. *hortte* **33,** 296–304; 10.21273/HORTTECH05207-23 (2023).

41. Gelernter, W. D., Stowell, L. J., Johnson, M. E. & Brown, C. D. Documenting Trends in Energy Use and Environmental Practices on US Golf Courses. *Crop Forage & Turfgrass Mgmt* **3,** 1–7; 10.2134/cftm2017.07.0044 (2017).

42. Walker, G. & Day, R. Fuel poverty as injustice: Integrating distribution, recognition and procedure in the struggle for affordable warmth. *Energy Policy* **49,** 69–75; 10.1016/j.enpol.2012.01.044 (2012).

43. Sovacool, B. K., Burke, M., Baker, L., Kotikalapudi, C. K. & Wlokas, H. New frontiers and conceptual frameworks for energy justice. *Energy Policy* **105,** 677–691; 10.1016/j.enpol.2017.03.005 (2017).

44. Lee, J.-R. & Kwon, K.-N. Popularity of Screen Golf in Korea and Its Sociocultural Meaning. *International journal of environmental research and public health* **18**; 10.3390/ijerph182413178 (2021).

45. Kengo Kamo. Former rural golf courses now sites for sea of solar power panels. Available at https://www.asahi.com/ajw/articles/14464677?utm_source=substack&utm_medium=email (2021).

46. Denholm, P. *et al.* Examining Supply-Side Options to Achieve 100% Clean Electricity by 2035. Available at https://www.nrel.gov/docs/fy22osti/81644.pdf (2022).

47. Schlemminger, M. *et al.* Land competition and its impact on decarbonized energy systems: A case study for Germany. *Energy Strategy Reviews* **55,** 101502; 10.1016/j.esr.2024.101502 (2024).

48. Golflux. Which Country Has the Most Golf Courses in the World? Available at https://www.golflux.com/countries-have-the-most-golf-courses-in-the-world/ (2023).

49. Pelser, T., Weinand, J. M., Kuckertz, P. & Stolten, D. ETHOS.REFLOW: Renewable Energy potentials workFLOW manager. Available at https://github.com/FZJ-IEK3-VSA/ethos.REFLOW (2024).

50. Spotify. Luigi. Available at https://github.com/spotify/luigi (2024).

51. Ryberg, D. S., Robinius M. & Stolten D. Evaluating Land Eligibility Constraints of Renewable Energy Sources in Europe. *Energies* **11**; 10.3390/en11051246 (2018).

52. Maier, R. *et al.* Potential of floating, parking, and agri photovoltaics in Germany **200,** 114500; 10.1016/j.rser.2024.114500 (2024).